\documentclass{article}

\def\lromn#1{\uppercase\expandafter{\romannumeral#1}}

\def\blist{\begin{list}{\setlength{\rightmargin}{\leftmargin}}}
\def\elist{\end{list}}
\def\be{\begin{equation}}
\def\ee{\end{equation}}
\def\bea{\begin{eqnarray}}
\def\eea{\end{eqnarray}}

\begin{document}

\begin{flushright}
TU/00/586
\end{flushright}

\begin{center}
\begin{large}

\renewcommand{\thefootnote}{\fnsymbol{footnote}}
\bf{Relic abundance of dark matter particles:
new formulation and new result of abundance calculation
}
\footnote[2]
{
Talk given at COSMO99, held at ICTP, Trieste, 
September 27-October 2, 1999.
To appear in the Proceedings (World Scientific).
}

\end{large}

\vspace{36pt}

\begin{large}
M. Yoshimura

Department of Physics, Tohoku University\\
Sendai 980-8578 Japan\\
\end{large}

\vspace{4cm}

{\bf ABSTRACT}

\end{center}

A new theoretical framework for computation of the relic abundance
of cold dark matter particles such as LSP is presented and some
generic features of new results are discussed.
The most important is a generalization of the Boltzmann
equation to include off-shell effects and its thermal average
over cosmic medium.
It is shown that at very low temperatures, much below the mass
of annihilating particles, equilibrium abundance is suppressed only
by powers of temperature instead of the exponential 
Boltzmann factor.
This change gives a larger relic abundance when  heavy
particles freeze out at these low temperatures.

\vspace{0.5cm} 
\section{
Introduction
}

\hspace*{0.5cm} 
The well-known problem in cosmology, the dark matter problem,
requires some sort of new heavy stable particles which have
decoupled from the rest of the universe at an early epoch.
The conventional estimate of the relic abundance \cite{review on lsp} 
of dark matter particles that have 
escaped from pair annihilation is based on 
the Boltzmann equation for the annihilating
particle for which the rate is
thermally averaged over the rest of light particles in
cosmic medium.
The freeze-out temperature of pair annihilation 
may roughly be estimated by equating the inverse pair annihilation
rate to the Hubble rate, and the relic abundance is determined
by the thermal number density at that freeze-out temperature
\cite{lee-weinberg}, which is exponentially suppressed at low
temperatures.

There are two possible issues as to whether this procedure 
gives a reliable result for the relic abundance.
First, the Boltzmann equation, although very appealing on
intuitive grounds, uses on-shell quantities such as
the cross section and the decay rate (more generally
S-matrix elements that relate a state at infinite past to that at
infinite future), while quantum mechanics
requires the more general transition amplitude at
finite times, or the Green's function.
The second source of concern is the use of the ideal gas form
of the distribution function or the occupation number
$1/(e^{E/T} - 1)$ at the freeze-out, 
which is very small at $E \approx M \gg T$
with $M$ the mass of the pair annihilating particle.
Both of these concerns are overcome and the Boltzmann approach
is indeed justified at high enough
freeze-out temperatures, as explicitly demonstrated in our works
explained here. But at very low freeze-out temperatures, say
$T < M/30$, this conventional procedure may well be questioned.

The key idea towards a more rigorous formulation is separation
of a small system, in this case the system of pair annihilating
particles, from the rest of a large thermal environment,
including the annihilation product.
By integrating out the environment part one may derive
a dynamical equation for the small system and may
obtain the quantum kinetic equation for the number density
of the annihilating particle.
If one works out the kinetic equation in a closed form and
analyzes the equilibrium number density, this could replace
the thermally averaged Boltzmann equation.
The problem thus raised is similar to the quantum Brownian
motion in thermal medium.

The report presented here is a short summary of our recent
works \cite{my-98}$-$\cite{my-99-2}.
Before we go to the pair annihilation model, we explain
a simpler toy model of coupled harmonic oscillators \cite{jmy-decay}, which
clarifies the essential point of the temperature power law.

\section{Linear open system}

\hspace*{0.5cm} 
We expect that 
the behavior of a small system immersed in thermal environment
is insensitive to the detailed modeling of the environment and
the form of its interaction to the system. 
Since the pioneering work of Feynman-Vernon \cite{feynman-vernon}  and 
Caldeira-Leggett \cite{caldeira-leggett} the standard model uses
an infinite set of harmonic oscillators (denoted by $Q(\omega )$)
for the environment and a bilinear interaction with a small system
(denoted by $q$);
\begin{eqnarray}
&& \hspace*{-0.5cm}
L_{Q} = \frac{1}{2}\, \int_{\omega _{c}}^{\infty }\,d\omega \,
\left( \,\dot{Q}^{2}(\omega )- \omega ^{2}\,Q^{2}(\omega )\,\right) \,,
\hspace{0.5cm} 
L_{{\rm int}} = -\, q\,\int_{\omega _{c}}^{\infty }\,d\omega \,
c(\omega )Q(\omega ) \,. 
\end{eqnarray}
We take for the system an unstable harmonic oscillator;
the potential $V(q) = \frac{1}{2}\, \omega _{0}^{2}q^{2}$
with $\omega _{0} > \omega _{c}$ the threshold energy.
This instability condition is imposed to mimic the unstable particle decay
and the pair annihilation process which has a threshold.

This standard toy model is exactly solvable, both by
the operator method and by the Feynmann-Vernon influence functional
method \cite{jmy-decay}.
We only quote the most essential part of the solution for our discussion;
the kinetic equation and the equilibrium occupation number
$f_{{\rm eq}}$ are given by
\begin{eqnarray}
&&
\frac{df}{dt} = -\,\Gamma \,(f - f_{{\rm eq}}) \,,
\hspace{0.5cm} 
f_{{\rm eq}} = \int_{\omega _{c}}^{\infty }\,d\omega \,
\frac{H(\omega )}{e^{\omega /T} - 1} \,, 
\\ &&
H(\omega ) \approx  
\frac{r(\omega )}{(\omega ^{2} - \omega _{0}^{2})^{2}
+ (\pi r(\omega ))^{2}} \,, \hspace{0.5cm} 
r(\omega ) = \frac{c^{2}(\omega )}{2\omega } \,.
\end{eqnarray}
This result is obtained in the Markovian approximation which
erases initial memory effects.
The occupation number is defined here by the thermal average
of the number operator of the system harmonic oscillator
over environment variables;
$f(t) \equiv \langle c^{\dag }(t)c(t) \rangle$.
We dropped a minor (in the weak coupling limit)
term of the frequency shift.
The relaxation rate is given by 
$\Gamma \approx \pi r(\omega _{0})/\omega _{0}$.

Let us look into the equilibrium occupation $f_{{\rm eq}}$ as 
a function of temperature $T$.
First of all, the physical meaning of the weight function $H(\omega )$
is that it is precisely the overlap probability of the original
dynamical variable $q$ and the eigen variable $\tilde{Q}(\omega )$
that diagonalizes the total Hamiltonian. Thus, a continuous integral
over $\omega $ appears for the equilibrium value.
Indeed, this formula for $f_{{\rm eq}}$ can also be understood by
the average of the occupation number using the
Botzmann-Gibbs ensemble of $e^{-\,H_{{\rm tot}}/T}$ 
with $H_{{\rm tot}}$ the total and not the free part of the Hamiltonian.

We consider a weak coupling region, 
$\Gamma \ll {\rm min.}\;(\omega _{0} \,, T)$
for the rest of our discussion.
Two important cases are worth of close examination.
The first is high temperature region in which
$T \geq \omega_{0}$.
In this case the Bose-Einstein distributiion function $1/(e^{\omega /T} - 1)$
hardly varies where the Breit-Wigner function $H(\omega )$ changes
over $\Gamma $, and one may replace $H(\omega )$ by the delta function 
$\delta (\omega - \omega _{0})$.
This gives
\( \:
f_{{\rm eq}} \approx 1/(e^{\omega _{0}/T} - 1) \,, 
\: \)
which is nothing but the on-shell form of the occupation number.
On the other hand, if the temperature is very low like
$T \ll \omega _{0}$, the region around $\omega \sim \omega _{0}$
gives a minor contribution in the continuous integral, and
the threshold region $\omega \sim \omega _{c}$ becomes more important.
Parametrizing this region by 
$r(\omega ) \approx c\,(\omega - \omega _{c})^{\alpha }$, one gets 
\( \:
f_{{\rm eq}} \approx 
O[1]\,\times \,c\,T^{\alpha + 1}/(\omega_{c}^{2} - \omega _{0}^{2})^{2} \,.
\: \)
The salient feature of this formula is appearance of the temperature
power law $\propto T^{\alpha + 1}$; 
it arises from the continuous $\omega $ integral outside
of the on-shell region $\omega \sim \omega _{0}$.
Even at intermediate temperatures the sum of two limiting formulas
is a good approximation;
\begin{equation}
f_{{\rm eq}} \approx  \frac{1}{e^{\omega _{0}/T} - 1} +
O[1]\,\times \,\frac{c\,T^{\alpha + 1}}
{(\omega_{c}^{2} - \omega _{0}^{2})^{2}}
 \,.
\end{equation}

Tha basic features of this simple model can be taken over to
the more complicated cases such as the unstable particle decay
and the pair annihilation to which we now turn to.

\section{Pair annihilation model}

\hspace*{0.5cm} 
We consider a simple boson pair annihilation model.
The heavy particle is denoted by the field $\varphi $ and
the light environment particle by $\chi $.
The interaction Lagrangian density that gives rise to the annihilation
process $\varphi \varphi \rightarrow \chi \chi $ is
\begin{eqnarray}
&&
{\cal L}_{{\rm int}} = -\,\frac{\lambda }{4}\,\varphi ^{2}\,\chi ^{2}
- \frac{\lambda_{\varphi}}{4!}\,\varphi ^{4}
- \frac{\lambda_{\chi}}{4!}\,\chi ^{4} + \delta {\cal L}
\,.
\end{eqnarray}
The last two terms ($\propto \lambda _{\varphi }$ or $\lambda _{\chi }$)
are introduced both for consistency
of renormalization and for thermalization of $\chi $ particles, 
hence the parameters are taken to satisfy
\( \:
|\lambda_{\varphi}|  \ll   \lambda^2 \ll 1
\,, \hspace{0.2cm} \; |\lambda_{\chi}| < 1 \,,
\: \)
but $|\lambda _{\chi }| \gg |\lambda |$.
The renormalization counter term $\delta {\cal L}$ is determined in
the usual way.

It is convenient for discussion of our approximation scheme
to employ the influence functional \cite{feynman-vernon}.
This quantity denoted by ${\cal F}[q(\tau ) \,, q'(\tau )]$ 
results after the environment integration for the density
matrix of the entire system.
Since the density matrix involves both the transition
amplitude and its conjugate, the path integral formula
resulting from the environment integration has a functional
dependence both on the path $q(\tau )$ and its conjugate path
$q'(\tau )$.
For the environment field $\chi $ we may take the Gaussian part
of the Lagrangian density since the $\lambda _{\chi }$ self-interaction
only serves for thermalization of the $\chi $ system.
With the thermal ensemble given by a Gaussian $\chi $ density matrix,
one can explicitly perform the $\chi $ path integration.
The result is a nonlinear quartic action;
\begin{eqnarray}
&& \hspace*{1cm}
{\cal F}[\varphi \,, \varphi '] =
\nonumber \\ && 
\exp [\,-\,\frac{1}{16}\,\int_{x_{0} > y_{0}}\,dx\,dy\,
\left( \,\xi _{2}(x)\,\alpha _{R}(x - y)\,\xi _{2}(y) + i\,
\xi _{2}(x)\,\alpha _{I}(x - y)\,X_{2}(y)\,\right)\,] \,, 
\nonumber 
\\ &&
\hspace*{1cm} 
X_{2}(x) \equiv  \varphi ^{2}(x) + \varphi '\,^{2}(x) 
\,, \hspace{0.5cm} 
\xi _{2}(x) \equiv \varphi ^{2}(x) - \varphi '\,^{2}(x) \,, 
\label{original influence functional} 
\\ && 
\hspace*{-0.2cm}
\alpha (x ) = \alpha _{R}(x ) + i\alpha _{I}(x )
= \lambda ^{2}\,\left( \,
{\rm tr}\;\left( T[\chi^{2}(x) \chi^{2}(0)\,
\rho _{\beta }^{(\chi )}]\right) 
- (\,{\rm tr}\; \chi ^{2}\rho _{\beta }^{(\chi)}\,)^{2}
\,\right) \,,
\label{2-point kernel} 
\end{eqnarray}
with $\rho _{\beta }^{(\chi )}$ the thermal density matrix for
the environment $\chi $.
The last equation for the kernel $\alpha $ means that
it is the real-time thermal Green's function.

One point is important before we further go on.
Since we deal with a non-equilibrium state for the $\varphi $ field,
time translational invariance is in general violated;
for the correlator 
\begin{equation}
\langle \varphi (\vec{x} \,, t_{1})\varphi (\vec{y} \,, t_{2}) \rangle
\neq 
\langle \varphi (\vec{x} \,, t_{1} - t_{2})\varphi (\vec{y} \,, 0) \rangle
\,,
\end{equation}
unlike the correlator in complete thermal equilibrium.
Thus, the correlator has dependence both on the relative and on
the central time.

It is crucial for further development to realize that
dependence on these two times is greatly different.
Variation in relative times is governed by the inverse of
the system frequency $1/\omega _{0}$ and is fast, while the central time
dependence is given by a much larger time scale, 
$1/(\lambda ^{2}\,\omega _{0})$ since without interaction
the system does not change.
It is thus reasonable to treat the central time dependence
adiabatically; one first regards the central time as
a constant and solves the dynamics with respect to the relative
time, and finally recovers the central time dependence at the end.

Another technical point for simplification is the
Hartree approximation.
The influence functional given above has four fields $\varphi $ at
two different times in the exponent. 
In a dilute system such as the pair annihilation in the expanding
universe it is a good approximation to ignore higher orders
of correlation. This is in accord with the spirit of the Hartree
approximation in which one replaces a pair of $\varphi $ fields
in the influence functional by the correlator yet to be
determined.
If one formally solves the model in the Hartree approximation, the
resulting correlator becomes a functional of the correlator itself, 
thus one arrives at a self-consistency equation for correlators.
This self-consistency equation is not very convenient for further
analysis and it is far better to derive from the self-consistency relation
time evolution equation, namely the quantum kinetic equation.

A convenient quantity for the kinetic equation is
a combination of Fourier transform of the correlator,
\begin{equation}
\sigma(k,t) = \int\, d^4(x - y)\, \langle \varphi(x)\,\varphi(y) \rangle
\,e^{ik\cdot (x - y)}
\,,
\end{equation}
with $t = (x_{0} + y_{0})/2$.
Since spatial homogeneity holds for the problem of our interest, 
this quantity $\sigma $ does not depend on the central spatial coordinate
$(\vec{x} + \vec{y})/2$.
We define the key quantity
\( \:
\tau ( k \,, t) = \sigma (-k \,, t)/\sigma _{-}(k\,, t) \,, 
\: \)
where $\sigma _{\pm }$ is even and odd parts.
With the slow variation as to the central time one has the kinetic
equation in the form \cite{my-99-2},
\begin{eqnarray}
 &~&\frac{d \tau(p,t)}{dt}
  =
  -\,\frac{\lambda^2}{2E_{p}}
  \int \frac{d^3k_{1}}{(2\pi)^32\omega_{1}}
  \int \frac{d^3k_{2}}{(2\pi)^32\omega_{2}}
  \int \frac{d^3p'}{(2\pi)^3}
  \int^\infty_0 \frac{dp'_0}{2\pi}\,
  (2\pi)^3\, H(p'\,, \infty )
  \nonumber \\
  &~&
  \left\{ \,
    \delta^{(4)}(p + p' - k_1 - k_2)
    \left[
      \tau \tau'(1 + f_1)(1 + f_2) - (1 + \tau )(1 + \tau')f_1f_2
    \right]
  \right.
  \nonumber \\
  &~& +
  2\delta^{(4)}(p + p' + k_1 - k_2)
  \left[
    \tau \tau'f_1(1 + f_2) - (1 + \tau )(1 + \tau')(1 + f_1)f_2
  \right]
  \nonumber \\
  &~& + 
  \delta^{(4)}(p + p' + k_1 + k_2)
  \left[
    \tau \tau'f_1f_2 - (1 + \tau )(1 + \tau')(1 + f_1)(1 + f_2)
  \right]
  \nonumber \\
  &~& +
  \delta^{(4)}(p - p' - k_1 - k_2)
  \left[
    \tau (1 + \tau')(1 + f_1)(1 +f_2) - (1 + \tau )\tau'f_1f_2
  \right]
  \nonumber \\
  &~&  +
  2\delta^{(4)}(p - p' + k_1 - k_2)
  \left[
    \tau (1 + \tau')f_1(1 + f_2) - (1 + \tau )\tau'(1 + f_1)f_2
  \right]
  \nonumber \\
  &~&
  \left. +
    \delta^{(4)}(p - p' + k_1 + k_2)
    \left[
      \tau (1 + \tau')f_1f_2 - (1 + \tau )\tau'(1 + f_1)(1 + f_2)
    \right]\,
  \right\} \,,
  \label{quantum kinetic eq} 
\end{eqnarray}
with 
\( \:
E_{p} = \sqrt{p^{2} + M^{2}} \,, \omega _{i} =
\sqrt{k_{i}^{2} + m^{2}} \,.
\: \)
In the right hand side $\tau = \tau (p\,, t) \,, 
\tau ' = \tau (p' \,, t)$.
The distribution function in the right hand side $f_{i}$ is given
by that of light $\chi $ particles in thermal equilibrium.
The weight function $H$ here obeys a self-consistency equation;
\begin{eqnarray} 
\hspace*{0.5cm} &&
  H(k \,, \infty) =
  \frac{r_-(k,\infty)}
  {(k^2 - M^2(T) - \Pi(k,\infty))^2 + (\pi r_-(k,\infty))^2}
\,, \\ 
  r_-(k,\infty) &=& 8\,\int^\infty_{-\infty} \frac{d^4k'}{(2\pi)^3}
  \, H(k' \,, \infty)\,
  r_\chi(k + k')\frac{e^{\beta{k_0}} - 1}
  {(e^{\beta(k_0 + {k'}_0)} - 1)(1 - e^{-\beta{k'}_0})} 
\,, \nonumber 
\\ 
r_{\chi }(\omega  \,, k) &=& 
\frac{\lambda ^{2}}{256\pi ^{2}}\left( \,
\sqrt{1 - \frac{4m^{2}}{\omega ^{2} - k^{2}}}\theta (\omega 
- \sqrt{k^{2} + 4m^{2}}) + \frac{2}{\beta k}
\ln \frac{1 - e^{-\beta \omega _{+}}}{1 - e^{-\beta |\omega _{-}|}}\,\right)
\,, \nonumber 
\label{two-body spectral} 
\end{eqnarray}
with $\beta = 1/T$ the inverse temperature and 
\( \:
\omega _{\pm } = \frac{\omega }{2} \pm \frac{k}{2}\,
\sqrt{1 - 4m^{2}/(\omega ^{2} - k^{2})} \,.
\: \)
We included the temperature dependent mass shift $M^{2}(T)$ of
order $\lambda $ and further $O[\lambda ^{2}]$ terms of
the proper self-energy $\Pi $.

The structural resemblance of this kinetic equation to the
Boltzmann equation is obvious and in this sense this equation
is its generalization.
There are a few important differences, however.
First, the basic quantity $\tau (p\,, t)$ is a function of 4-momentum
$p$ unlike the distribution function 
$f(\vec{p} \,, t)$ in the Boltzmann
equation, which only depends on the 3-momentum.
Related to this is that even the processes not allowed by the energy-momentum
conservation on the mass shell contribute to the collision term
in the right hand side, for example $1 \leftrightarrow 3$ process
such as $\varphi \leftrightarrow \varphi \chi \chi $.

Obviously, the equilibrium solution for which the right hand side
vanishes is
\( \:
\tau (p \,, \infty ) = 1/(e^{\beta p_{0}} - 1) \,.
\: \)
We assume that this solution is unique and no other solution exists.
One further determines the equilibrium correlator by
\begin{equation}
\langle \varphi(x) \varphi(y) \rangle_{{\rm eq}} =
\int\,\frac{d^4 k}{(2\pi)^3}\,\frac{1}{1 - e^{-\,\beta k_{0}}}\,
H(k \,, \infty) \,e^{-i\,k\cdot(x - y)} \,.
\label{asymp thermal correlator} 
\end{equation}
It can be shown that to $O[\lambda ^{2}]$ this correlator coincides with
that given by the thermal field theory \cite{my-99-1}.

The weight function $H$ or $r_{-}$ obeys the self-consistency
equation. We analyze this and the kinetic equation perturbatively.
In the zero coupling limit
\( \:
H(p\,, \infty ) \approx \delta (p_{0} - E_{p})/2E_{p} \,
\: \)
for $p_{0} > 0$,
and our kinetic equation reduces to the usual Boltzmann equation.
To include the next $O[\lambda ^{2}]$ order the weight function 
$H(k\,, \infty )$ is calculated by using the integral form of
$r_{-}(k\,, \infty )$ in which  $H$ is replaced by the delta function.
The resulting kinetic equation with $H(k\,, \infty )$ to this order
and its associated equilibrium formula gives our basic results.

Once the kinetic equation is solved, one can compute physical
quantities using the function $\tau $.
The $\varphi$ energy density is thus calculated taking
the coincident time limit of two-point correlators, with due consideration
of renormalization of composite operators.
Thus, 
\begin{eqnarray}
  \langle \cal{H}_{\varphi} \rangle
  &= &
  \langle
  \frac{1}{2}\dot{\varphi}^2
  + \frac{1}{2}(\nabla{\varphi})^2
  + \frac{M^2}{2}\varphi^2 - ({\rm counter \; terms})
  \rangle
  \\
  &\simeq &
  \int\frac{d^3p}{(2\pi)^3}\int^\infty_0 dp_0\,
  (\,p_0^2 + E_p^2\,)H(p\,, \infty )\tau(p,t)
  \nonumber \\
  &+& \frac{\lambda ^{2}}{32\pi ^{2}}\int\,\frac{d^{3}kd^{3}k'}
  {(2\pi )^{6}4\omega \omega '}(k^{2} + k'\,^{2})f_{k}f_{k'}
  \int_{0}^{\infty }\,dp\,\frac{p^{2}}{E_{p}} 
  \nonumber \\
  &\simeq&
  \int\frac{d^3p}{(2\pi)^3}\,
  E_p\, \tau(E_p,\vec{p}\,,t)
  + c\,\lambda^2\frac{T^6}{M^2} \,, \label{energy density} 
\end{eqnarray}
with
\( \:
  c = \frac{1}{69120} \sim 1.4\,\times 10^{-5} \,.
\: \)
(The $\chi $ mass $m$ was taken to vanish for this calculation.)
In the last step we replaced the spectral weight $H$ by the
delta function.
The temperature power term ($\propto T^{6}$) thus arises, as expected.
The first term for the energy density involves the $\tau $
function on the mass shell, hence is Boltzmann suppressed at
low temperatures.

The temperature power $T^{6}$ is understood as follows.
Since the operators for ${\cal H}_{\varphi }$ have mass dimension 2 or 4,
terms of order $T^{2}$ and $T^{4}$ are divergent and they are
cancelled by counter terms.
The remaining finite term is of order $T^{6}$, as given above. 
Precisely the same temperature power term was also derived
by using the technique of thermal field theory \cite{my-99-1},
which is hardly surprising since our equilibrium result is given
by the Boltzmann-Gibbs ensemble.

\section{Cosmological evolution}

\hspace*{0.5cm} 
We find it reasonable to define the $\varphi $ number density
by $n_{\varphi } = \rho _{\varphi } / M$ at low temperatures of
$T \ll M$.
The time evolution equation for the number density
in the expanding universe is given by
\begin{equation}
\frac{dn_{\varphi }}{dt} + 3\,H \,n_{\varphi } =
-\,\langle \sigma v \rangle\,(\,n_{\varphi }^{2} - n_{{\rm eq}}^{2}\,)
\,, 
\label{time evol for n} 
\end{equation}
with 
$H$ the Hubble rate,
\( \:
\langle \sigma v \rangle \approx \lambda ^{2}/(32\pi \,M^{2})
 \,,
\: \)
and
\begin{equation}
n_{{\rm eq}} \approx (\frac{MT}{2\pi })^{3/2}\,e^{-\,M/T} +
c\,\lambda ^{2}\,\frac{T^{6}}{M^{3}}\,.
\end{equation}
The usual time-temperature relation $t \propto T^{-2}$ may be used.
Time evolution calculated from eq.(\ref{time evol for n}) supports
the picture of the sudden freeze-out as in the Lee-Weinberg analysis
\cite{lee-weinberg}.
The major difference here is the new equilibrium abundance $n_{{\rm eq}}$.
When the freeze-out temperature is high enough, our result justifies
the usual thermally averaged Boltzmann approach.
On the other hand, 
when the freeze-out occurs at which the temperature power term dominates,
the final relic abundance is enhanced over the Boltzmann abundance.

For numerical results I refer to our paper \cite{my-99-1}.
Our new effect tends to show up for a larger coupling and a smaller
mass.
Since the new effect gives an additional positive contribution to 
the energy integral,
the relic density is always enhanced over the conventional
result.
Thus the allowed parameter region in the model parameter space
gets always shrunk by our result.

\end{document}